# OSI-flex: Optimization-Based Shearing Interferometry for Joint Phase and Shear Estimation Using a Flexible Open-Source Framework

Julianna Winnik[1,*], Damian Suski[2], Matyáš Heto[1], Małgorzata Lenarnik[3,4], Michał Ziemczonok[1], Maciej Trusiak[1] and Piotr Zdańkowski[1,*]

[1] Institute of Micromechanics and Photonics, Faculty of Mechatronics, Warsaw University of Technology, Warsaw, Poland
[2] Institute of Automatic Control and Robotics, Faculty of Mechatronics, Warsaw University of Technology, Warsaw, Poland
[3] Department of Pathology, Maria Sklodowska-Curie National Research Institute of Oncology, Warsaw, Poland
[4] Department of Gastroenterology, Hepatology and Clinical Oncology, Centre of Postgraduate Medical Education, Warsaw, Poland

*Authors to whom any correspondence should be addressed.

E-mail: julianna.winnik@pw.edu.pl, piotr.zdankowski@pw.edu.pl

## Abstract

Shearing interferometry is a common-path quantitative phase imaging technique in which an object beam interferes with a laterally shifted replica of itself, providing high temporal stability, reduced sensitivity to environmental noise, compact design, and compatibility with partially coherent illumination that suppresses coherence-related artifacts. Its principal limitation, however, is that it yields only sheared phase-difference measurements rather than the absolute phase, thereby requiring additional reconstruction step. In this work, we introduce OSI-flex, a flexible, open-source computational framework for quantitative phase reconstruction from sheared phase-difference measurements. The method leverages modern machine learning tools, namely automatic differentiation and the advanced ADAM (Adaptive Moment Estimation) optimizer. The method simultaneously refines the phase distribution and estimates the shear values, enabling it to adapt to experimental conditions where the shear cannot be precisely determined. Because defining shear value is inherently difficult in most systems, yet crucial for effective phase reconstruction, this joint optimization leads to robust and reliable phase retrieval. The proposed OSI-flex framework is highly versatile, supporting arbitrary numbers, magnitudes, and orientations of shear vectors. While optimal reconstruction is achieved with two orthogonal shear directions, the inclusion of regularization—specifically total variation minimization and sign constraint—enables OSI-flex to remain effective with nonorthogonal or even single-shear measurements. Moreover, the algorithm accommodates a wide range of shear magnitudes, from subpixel shifts (differential configuration) to several dozen pixels (semi-total shear configuration). Validation with simulations and experimental data confirms quantitative accuracy on calibrated phase objects and demonstrates robustness with 3D-printed cell phantom and follicular thyroid cells.

## 1. Introduction

Quantitative phase imaging (QPI) [1,2] refers to a set of label-free optical techniques that map the object-related phase shift of light, providing high-resolution measurements of optical path length variations caused by changes in refractive index and thickness. This makes QPI a powerful tool for studying cellular structures, tissue morphology, and other delicate specimens without the need for staining or labeling.

Among various QPI architectures, common-path interferometric systems [3–8] offer several key advantages: enhanced temporal stability due to identical optical paths for the reference and sample beams, reduced sensitivity to environmental perturbations, a compact optical layout and related possibility of easy integration with existing microscopes as add-on modules, lower susceptibility to system-induced aberrations, and—critically—the ability to operate with partially coherent illumination, which minimizes coherence-related artifacts and noise.

One implementation of common-path QPI is shearing interferometry (SI), especially lateral SI [9], in which the object beam interferes with a laterally shifted copy of itself. SI is often implemented in a total shear configuration [5,10–14] where the shear is sufficiently large to ensure that the object beam interferes with an object-free region of its replica. While effective for sparse samples, this approach fails for confluent or spatially extended specimens. To address this limitation, point diffraction interferometry [6], diffraction phase microscopy (DPM) [8,15], and other related setups [16] have been proposed, where the object beam copy is cleared with the use of a pinhole. However, these systems are complex to align and sensitive to mechanical instabilities, which limits their practical use in dynamic or high-throughput settings.

An alternative approach is to apply SI with the shear magnitude minimized below the system's spatial resolution. In this configuration, the resulting fringe pattern encodes the spatial derivative of the object phase along the shear direction. SI can be implemented in various ways, for example using a Wollaston prism in Nomarski's differential interference contrast microscopy [17], shear plate [18–20], interferometers such as Michelson [21–23] or Sagnac [24], 2D diffraction gratings in quadriwave lateral shearing interferometry [25–29], or using polarization diffraction gratings [30–32].

More broadly, the class of differential phase contrast architectures within the QPI family extends beyond SI approach. Examples include Shack–Hartmann sensor-based measurments [33], differential phase contrast [34], gradient light interference microscopy [35,36], gradient retardance optical microscopy [37], metasurface enabled microscopy [38,39], oblique plane back illuminated microscopy [40,41], and optical differentiation using spatially variable amplitude filters [42]. These approaches provide alternative means of retrieving phase derivatives, often along multiple directions.

For certain applications, such as strain analysis [43,44], slope measurement [45], lateral aberrations evaluation [46], and 3D refractive index gradient imaging [31], measuring the phase derivative alone may be sufficient. In most cases, however, the derivative (or a set of derivatives along different directions) serves merely as an intermediate step, providing input for a numerical integration procedure.

A variety of computational integration methods have been used in SI. The most straightforward approach is direct numerical integration [47–49], which is simple and fast. However, it requires two small orthogonal shears and is prone to noise accumulation and error propagation. Another widely used technique is Fourier transform-based reconstruction [7,50–53], which typically relies on two small orthogonal shears but has also been generalized to accommodate orthogonal finite shears magnitudes [26]. A third approach is the iterative method known as alternating projections [30,54]. While effective in some cases, this procedure is heuristic in nature and often demands a large number of phase-difference measurements—sometimes as many as 24 [55].

Phase integration can also be addressed using optimization-based approaches. For smooth phase distributions, as in wavefront sensing, so called modal methods can be applied. In this approach, coefficients of the modal functions are estimated using least-squares fitting. The popular modal functions are polynomials [49,56,57], including most relevant Zernike polynomials [19,20,56,58,59]. While this strategy effectively suppresses errors caused by random noise, it also imposes a limitation on the object being investigated to ensure the wavefront can be adequately represented by the modal functions.

Another group of so-called zonal algorithms minimizes reconstruction errors locally. This approach generally yields higher detail but is more computationally demanding and more susceptible to noise. This group includes least-squares integration with Southwell geometry [60], where reconstructed wavefronts coincide with local slope measurements, as well as related improved algorithms that aim at increased accuracy [61,62] and reduced computation time [63]. Another approach employs spline-based least-squares integration [33,64]. Phase restoration can also be performed using partial differential equation methods [65], which reconstruct the wavefront by solving the linear system of equations that relates the reconstructed phase and its bidirectional difference. Notably, although the aforementioned zonal methods are in principle applicable to arbitrary phase maps, they have mostly been applied to smooth, slowly varying distributions [33,60–65] rather than to classical imaging cases.

Moreover, exisiting alogirthms for object phase reconstrcution in SI impose various strict restritions on the shear vectors, such as requiring a pair of perfectly orthogonal *x*- and *y*- oriened shears[21,25,26,49,50,52,53,61,64,66], multiple SI measurements in diversfied directions [30], small shears magnitudes [21,52] or even shears of exactly one-pixel magnitudes [22,23,65].

Additionally, it is important to note that accurate knowledge of the shear values is a critical requirement for phase estimation from the SI data [19,30]. Even small errors in these parameters can substantially compromise the accuracy [19] and resolution [30] of phase reconstruction. Shear values can be estimated with the use of a calibration object. They can also be calculated from correlation maps with a precision of up to 0.5 pixels [30], which may be insufficient in high-resolution and high-fidelity reconstruction scenarios.

In this paper, we present a flexible, open-source optimization-based computational framework for SI, called OSI-flex, which enables quantitative phase reconstruction from sheared phase differences. Our method leverages well-established and efficient machine learning tools, namely automatic differentiation and the ADAM (Adaptive Moment Estimation) optimization algorithm. The core of the approach is a joint optimization of the object phase distribution and shear vectors, making it adaptable to experimental setups with imprecise shear parameters. By introducing a joint optimization strategy, we improved robustness and overall reconstruction accuracy.

The proposed OSI-flex algorithm is flexible with respect to the number, magnitudes, and orientations of applied shears. It accommodates shear magnitudes ranging from subpixel values to several dozen pixels. Incorporating regularization into the optimization framework improves the conditioning of the inverse problem, thereby ensuring applicability to nonorthogonal shears and enabling information-bearing reconstruction even under the degenerate case of single-shear measurements.

The algorithm is validated through simulations, demonstrating its versatility, and confirmed on experimental data. Quantitative accuracy is verified using calibrated phase objects, while its utility and robustness are further demonstrated with a 3D-printed phantom with dense cell confluence and a thyroid smear sample. The algorithm code is publicly available as open-source software.

## 2. Methods

### *2.1 Experimental system*

Figure 1 shows the schematic of our common-path SI system, which was previously developed in our group [14,31]. In the system a collimated LED beam illuminates the sample and then passes through the objective lens (OL). The object beam is subsequently collimated by the tube lens (TL) before entering the shearing module (SM), which produces two laterally sheared replicas of the object beam.

The SM consists of two identical polarization diffraction gratings (PGs) placed parallel and in proximity. When the linearly polarized object beam strikes the first PG, it is split into two circularly polarized diffraction orders of opposite handedness. These two replicas then pass through the second PG. Since each beam has a single polarization component, the second PG does not split them further but only deflects them. Consequently, at the output of the SM we obtain two object beams propagating in parallel directions, with a precisely controlled shear magnitude and orientation. The shear magnitude is adjusted by changing the inter-grating distance, while the shear orientation is controlled by rotating the entire SM.

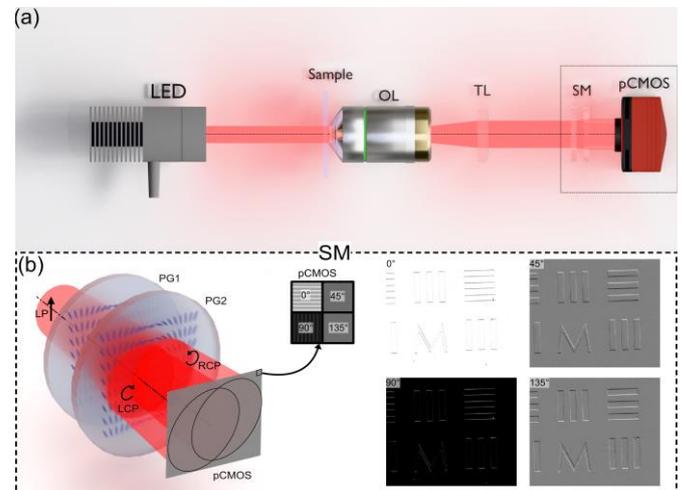

**Figure 1.** (a) Schematic of the polarization gratings-based SI system: OL – objective lens, TL – tube lens, SM – shearing module, pCMOS – polarization camera; (b) schematics of the SM: PG – polarization grating, LP – linear polarization, LCP/RCP – left and right handed circular polarizations, respectively, 0°, 45°, 90° and 135° - orientation of the pCMOS polarizing filter and corresponding intensity images for 0, $\pi/2$, $\pi$, $3\pi/2$ phase shifts, respectively.

The resulting sheared beams interfere, and the resultant fringe pattern is recorded by a polarization-resolved camera (pCMOS). This detector incorporates a micro-polarizer array with four linear polarizers at different orientations, enabling simultaneous acquisition of four interferograms with relative phase shifts of π/2. Since rotating the linear polarizer induces a phase shift between the two beams of opposite circular polarization, this setup provides four phase-shifted interference patterns in a single shot, thereby enabling reliable phase retrieval using the phase-shifting method [67].

*2.2 Interpretation of SI data*

The fringe patterns generated with SI are modulated by the difference between the object wave phase $\varphi$ and its transversely shifted copy:

$$\Delta\varphi(x, y; \boldsymbol{d}) = \varphi(x + d_x, y + d_y) - \varphi(x, y), \quad (1)$$

where $\boldsymbol{d}=(d_x,d_y)$ is the shear vector. The phase difference $\Delta\varphi$ can be retrieved from the fringe pattern using various phase-retrieval methods, e.g., phase shifting [67], the Fourier transform [68] or the Hilbert transform method [69].

For small (sub-resolution) shears, the measured distribution $\Delta\varphi$ can be interpreted as a finite-difference approximation of the phase derivative in the shear direction. In certain applications, $\Delta\varphi$ already contains useful information in its raw form. However, in most cases, $\Delta\varphi$ does not represent the final measurement, and a numerical algorithm is required to recover the true object phase distribution, $\varphi$. This usually entails capturing multiple phase-difference measurements, typically two with orthogonal shear orientations.

## 3. Results

*3.1. OSI-flex algorithm*

*3.1.1 Joint phase and shear estimation*

To address the problem of reconstructing the object phase from a set of $N$ measured phase differences $\Delta\varphi_n^M, n = 1, \ldots, N$ an optimization-based approach is adopted. The goal is to find a phase estimate $\bar{\varphi}$ that minimizes the cost function:

$$\mathcal{E}(\varphi; \boldsymbol{d}, \Delta\varphi^M) = \mathcal{D}(\varphi; \boldsymbol{d}, \Delta\varphi^M) + \mathcal{R}(\varphi), \quad (2)$$

where the data fidelity term $\mathcal{D}$ measures the discrepancy between the estimated and measured phase differences, and $\mathcal{R}$ serves as a regularization term, which improves well-posedness of the inverse problem and stabilizes the optimization procedure.

The data fidelity term is defined here as the $L_2$ loss, i.e., the mean squared error (MSE):

$$\mathcal{D}(\varphi; \Delta\varphi^M, \boldsymbol{d}) = \frac{1}{N}\sum_{n=1}^{N}\sum_{i=1}^{I}\sum_{j=1}^{J}\left[\tilde{\Delta}\varphi_n(i,j; \boldsymbol{d_n}) - \Delta\varphi_n^M(i,j)\right]^2, \quad (3)$$

where $(i,j)$ are the pixel indices in the discretized phase-difference maps (after phase unwrapping), and $n$ denotes the index of the measured phase difference map corresponding to the shear vector $\boldsymbol{d_n}$. For non-integer shears $d_x$, $d_y$ the phase difference estimates $\tilde{\Delta}\varphi(d_x, d_y)$ are computed using bilinear interpolation:

$$\begin{aligned}\tilde{\Delta}\varphi(d_x, d_y) &= w_x \cdot w_y \cdot \Delta\varphi(\lfloor d_x \rfloor, \lfloor d_y \rfloor) + \\ &\quad w_x \cdot (1 - w_y) \cdot \Delta\varphi(\lfloor d_x \rfloor, \lceil d_y \rceil) + \\ &\quad (1 - w_x) \cdot w_y \cdot \Delta\varphi(\lceil d_x \rceil, \lfloor d_y \rfloor) + \\ &\quad (1 - w_x) \cdot (1 - w_y) \cdot \Delta\varphi(\lceil d_x \rceil, \lceil d_y \rceil),\end{aligned} \quad (4)$$

where $\lfloor a \rfloor, \lceil a \rceil$ denote the floor and ceiling functions, $\Delta\varphi(\lfloor d_x \rfloor, \lfloor d_y \rfloor)$ and respective terms are phase differences calculated for integer shears closest to $d_x$, $d_y$ and the interpolation weights are $w_x = d_x - \lfloor d_x \rfloor$ and $w_y = d_y - \lfloor d_y \rfloor$. The interpolation function calculates the estimated phase difference for non-integer shears based on four phase differences with integer shears, which are obtained simply by shifting the phase image by some number of pixels. It should be noted that the definition (4) makes the interpolation function piecewise-differentiable with respect to shears, as when e.g. $\lfloor d_x \rfloor$ changes its value, the phase images taken for calculations are shifted by a different number of pixels.

The regularization term $\mathcal{R}$ consist of two weighted factors, i.e., total variation (TV) and sign penalty (SP) terms:

$$\mathcal{R}(\varphi) = \alpha_{TV}TV(\varphi) + \alpha_{SP}SP(\varphi). \quad (5)$$

The TV minimization:

$$TV(\varphi) = \sum_{i=1}^{I-1}\sum_{j=1}^{J-1}\sqrt{[\varphi(i+1,j) - \varphi(i,j)]^2 + [\varphi(i,j+1) - \varphi(i,j)]^2}, \quad (6)$$

promotes the piecewise constant solution and decreases the noise. The SP term penalizes violations of the nonnegativity/nonpositivity constraint:

$$SP(\varphi) = \sum_{i=1}^{I}\sum_{j=1}^{J} max\,(s \cdot \varphi(i,j), 0), \quad (7)$$

where

$$s = \begin{cases} -1 & for\ nonnegativity \\ +1 & for\ nonpositivity \end{cases}. \quad (8)$$

Again, it should be noted that defining the regularization term with the help of square root and maximum functions, makes it a piecewise differentiable function with respect to phase.

Our approach was designed with versatility in mind. Unlike existing methods, it does not restrict the number of available phase-difference maps, nor does it impose constraints on the magnitude nor orientation of the shear vectors. The algorithm can handle a wide range of shear magnitudes, from sub-pixel displacements to several dozen pixels. This flexibility is important, since practical hardware limitations often prevent an interferometer from producing very small shears. For example, in our system (Fig. 1), the minimal shear magnitude is about 2–3 pixels, and it is limited by the minimum spacing imposed by the grating housings. The performance of the algorithm under different shear magnitudes and orientations is analysed in Section 3.1.

Furthermore, our method explicitly addresses the common issue that the exact shear value is often not known with high precision—a factor that can strongly affect the accuracy of phase retrieval. To mitigate this, we treat the shear vectors **d$_n$** as the optimization variables and jointly solve the optimization problem for $\varphi$ and **d$_n$**.

A third factor contributing to the versatility of our algorithm is its reliance on established, well-tested, open-source software for numerical optimization. Specifically, we use the built-in ADAM optimizer [70] from the TensorFlow library, combined with automatic differentiation to compute gradients of the cost function with respect to the optimization variables: $\varphi$, **d$_n$**. This approach offers several advantages. First, it leverages an optimally implemented and widely validated optimization routine. Second, it provides flexibility, since the underlying mathematical model can be easily modified—for example, switching from a lateral to a radial [71] or azimuthal [72] SI configuration requires only a small change in the forward model. Finally, the use of TensorFlow 2's automatic differentiation is particularly beneficial: by constructing a computational graph, the framework simplifies derivative calculations and ensures that the computed gradients are fully consistent with the numerical implementation of the forward model.

The OSI-flex algorithm is publicly available at [73].

*3.1.2 Adam optimization*

To numerically minimize the cost function (Eq. (2)) with respect to phase and shears, the TensorFlow [74,75] implementation of the ADAM algorithm is used. In ADAM, the update rule for the decision variable $a$ ($\varphi(i,j)$ or $d_{[x;y],n}$ in our case) is [70]:

$$a^k = a^{k-1} - \eta \frac{\widehat{m}_a^k}{\sqrt{\hat{v}_a^k + \epsilon}}, \quad (9)$$

where $k$ is the iteration number, $\epsilon$ is a small scalar used to prevent division by 0 and $\eta$ is the global learning rate (for shears estimation the learning rate is additionally multiplied by the factor $\gamma$). The estimates of the first $\widehat{m}_a^k$ and second $\hat{v}_a^k$ gradient moment are updated according to the rules [70]:

$$m_a^k = \beta_1 \cdot m_a^{k-1} + (1-\beta_1) \cdot \nabla_a \mathcal{E}^{k-1}, \quad (10)$$

$$\widehat{m}_a^k = \frac{m_a^k}{1-\beta_1^k}, \quad (11)$$

$$v_a^k = \beta_2 \cdot v_a^{k-1} + (1-\beta_2) \cdot (\nabla_a \mathcal{E})^2, \quad (12)$$

$$\hat{v}_a^k = \frac{v_a^k}{1-\beta_2^k}, \quad (13)$$

where $0 < \beta_1, \beta_2 < 1$ are the forgetting coefficients of running averages expressed with Eqs. (10) and (12), and $\nabla_a \mathcal{E}^k$ is the partial derivative of the cost function with respect to the variable $a$ calculated at the $k$-th iterate. The formulas (10) and (12) express the running averages of the gradient and element-wise squared gradient. The additional updates (11) and (13) are introduced to diminish the bias of the running averages towards zero for initial iterations.

ADAM belongs to the class of algorithms originally developed to perform the training of deep neural networks (DNNs). For that reason, ADAM possesses the properties that are beneficial also for OSI-flex. First, the DNNs training is a large-scale optimization task in terms of the number of decision variables $a$. The discussed phase reconstruction task is also a large-scale task, as the number of decision variables is the image size in pixels (plus double the number of shear images). ADAM, a prominent example of DNN training algorithms, can handle a large number of decision variables primarily due to the structure of its update equations (Eqs. 9–13), which are computed independently for each variable. This independence enables parallelization of the updates and efficient utilization of parallel processing units, such as CPUs or GPUs

Another advantage of DNNs training algorithms is their ability to handle piecewise differentiable functions [76] (e.g., ReLU activation functions), which is also the case of our optimziation problem. Potentially faster-converging algorithms (e.g., conjugate gradient method [77]) heavily rely on the differentiability assumptions thus are not well suited to our problem.

Calculating the first (Eq. (10)) and second (Eq. (12)) gradient momenta allows for adjusting the update step size on the basis of approximate second order (curvature) behaviour of the cost function along the solution enhancement path [76]. Although ADAM lacks the theoretical convergence property [78], in practice it exhibits the fastest convergence among the most popular DNNs training algorithms [70].

ADAM, belonging to the gradient-with-momentum family of algorithms, provides a straighforward, parallelizable and easy to track scheme of computations. That stays in contrast to more complicated algorithms such as the proximal gradient algorithm [79], which requires solving subproblems at each iteration and using advanced heuristics to provide the algorithm's convergence.

Finally, leveraging the well-established implementation of ADAM, makes our OSI-Flex implementation stable, efficient and error-prone.

## 3.2 Simulation study

We evaluated OSI-flex with simulations using the ground-truth phase distribution in Fig. 2. Phase differences were generated via bilinear interpolation. The study assessed the influenced of shears number, magnitude, and orientation, as well as the impact of shear errors.

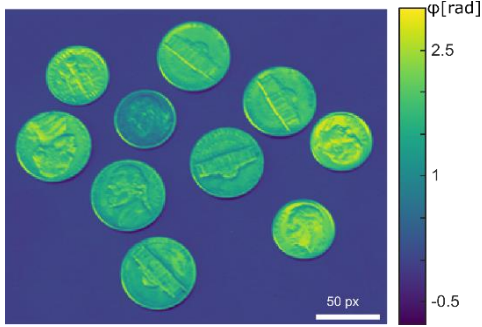

**Figure 2.** Ground-truth phase distribution used for simulated numerical studies.

### 3.2.1 Effect of shear magnitude

In SI systems, we usually aim for small shear magnitudes to ensure that finite differences $\Delta\varphi_n$ provide a good approximation of the phase derivative. In certain optical systems, however, generating very small shears can be challenging. Our algorithm is flexible in this matter, it is capable of handling both sub-pixel and large shear values. To evaluate performance across scales, three shear non-integer magnitudes were considered: small (0.8 pixel), medium (5.3 pixels), and large (50.3 pixels). In all cases, we considered a pair of orthogonal shearing directions. The simulated data is displayed in Fig. 3.

The reconstruction was performed using the following parameters: $\eta=0.1$, $\gamma=0$ (no shear fine-tuning), $\alpha_{TV}=1e^{-8}$, $\alpha_{SP}=1e^{-8}$. The results are shown in Fig. 4 (retrevied phases and error maps) and visualized in *Visualization 1* (retrevied phases across iterations). Reconstruction was carried out over 800 iterations, with the first 400 iterations emphasizing high spatial frequencies and the subsequent 400 iterations focusing on retrieval of low frequencies (see *Visualization 1*).

The optimization monitoring plots—(a) cost function, (b) structural similarity index (SSIM), and (c) root-mean-square error—are shown in Fig. 5. The quality metrics after 400 and 800 iterations are also presented in Tab. 1. As seen in Fig. 4, after 800 iterations, the reconstruction quality was high and generally consistent across all cases. Notably, for medium and large shears, convergence was achieved earlier (see *Visualization 1*). This is advantageous, as realizing very small shear can be challenging in some optical systems. Nevertheless, larger shear values introduce potential drawbacks, such as border artifacts and reduced fringe contrast due to the limited coherence of the light source. As a result, the medium shear case seems to provide the most favorable compromise between performance and experimental feasibility.

The 800 iterations of phase retrieval for the coin image (454×454 pixels) required approximately 31 s on a CPU (Intel Core i7-7700HQ, 2.80 GHz, 32 GB RAM).

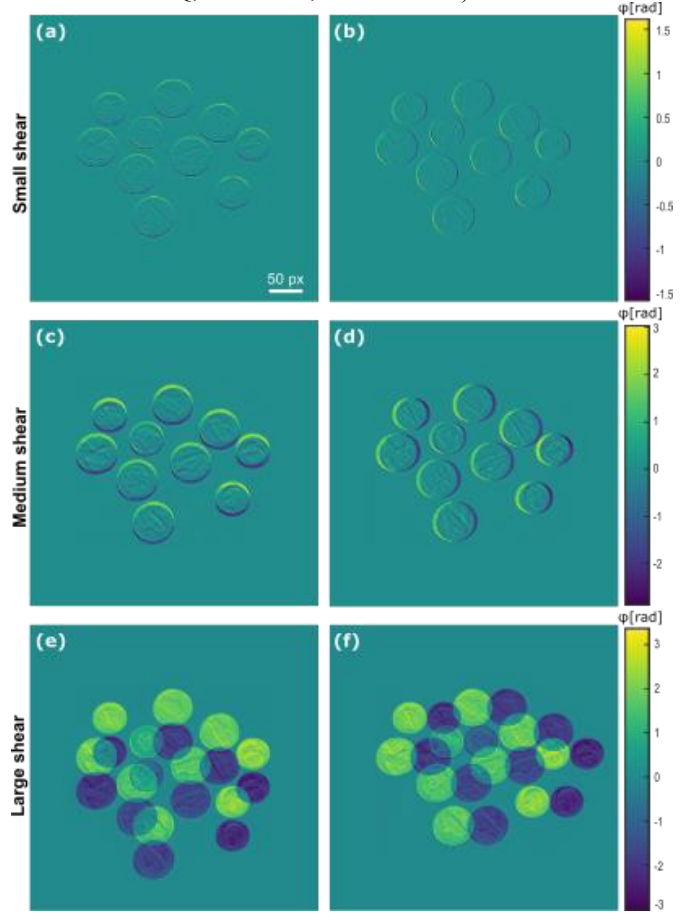

**Figure 3.** Shearing phase differences for a pair of orthogonal shears of varying magnitudes: (a), (b) small; (c), (d) medium; (e), (f) large.

**Table 1.** Quality metrics of reconstructed object phases after the 400[th] and 800[th] iterations for various shear magnitudes.

| Shear magnitude | | Small | Medium | Large |
|---|---|---|---|---|
| 400 iterations | SSIM | 0.6724 | 0.6696 | 0.9924 |
| | RMSE | 0.3280 | 0.1231 | 0.0058 |
| 800 iterations | SSIM | 0.9544 | 0.9975 | 0.9931 |
| | RMSE | 0.0257 | 0.0035 | 0.0055 |

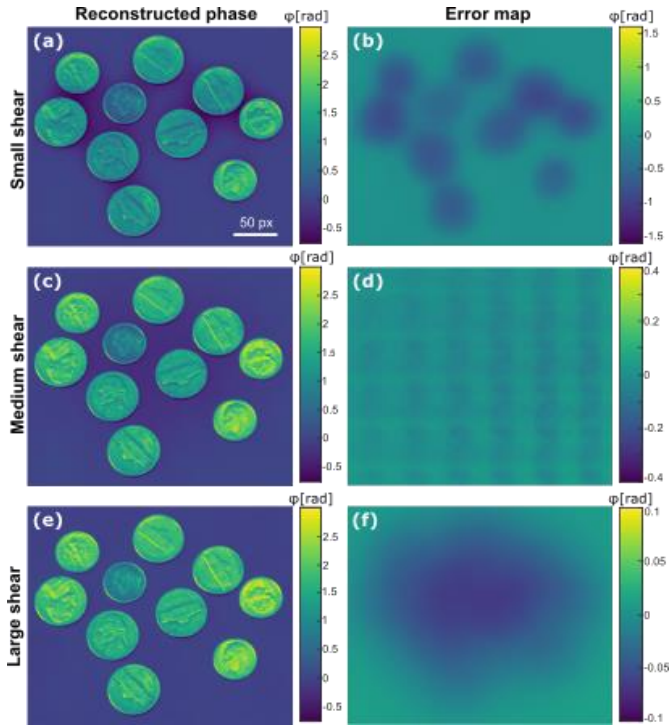

**Figure 4.** Reconstructed object phases and corresponding error maps for (a), (b) small, (c), (d) medium, and (e) (f) large shear magnitudes. *Visualization 1* shows the retrieved phases across iterations.

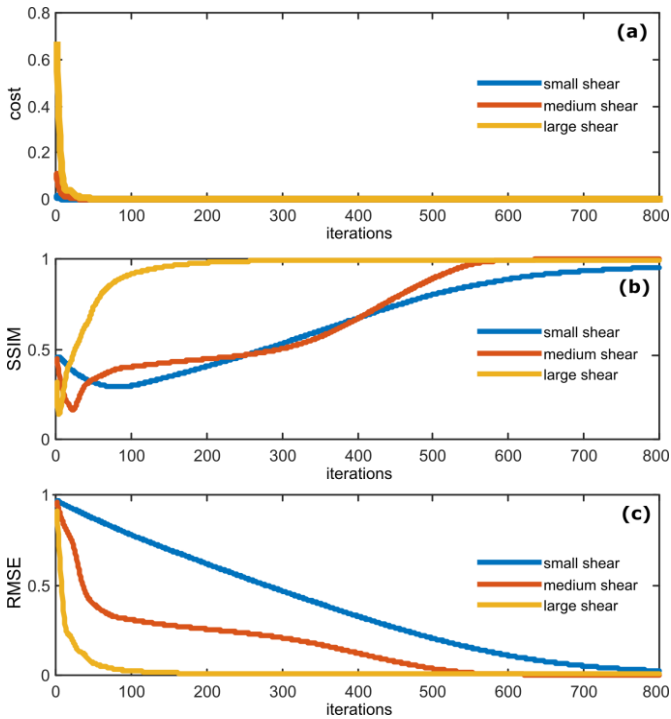

**Figure 5.** Optimization monitoring curves showing (a) cost and quality metrics ((b) SSIM and (c) RMSE) across iterations.

### 3.2.2 Unequal shear magnitudes

This section examines the case of two orthogonal shears with uneven magnitudes, i.e., 5.3 pixels in the vertical direction and 0.8 pixels in the horizontal direction. From previous tests, we observed that improvements beyond 400 iterations are moderate; therefore, in this study, we limit the computation to 400 iterations, corresponding to approximately 15.5 s of processing time.

In the previous section, the regularization coefficients were set to minimal values solely to stabilize the training against numerical errors. Under more challenging conditions—such as the presence of noise, data errors, or ill-posed problems—increasing the regularization can be beneficial. Here, we doubled the coefficient of the SP regularizer; thus, the applied parameters are: $\eta=0.1$, $\gamma=0$, $\alpha_{TV}=1e^{-8}$, $\alpha_{SP}=2e^{-8}$.

Figure 6 shows the retrieved phase and error maps. OSI-flex achieved high-quality restoration under uneven shears, with SSIM = 0.9108 and RMSE = 0.0515.

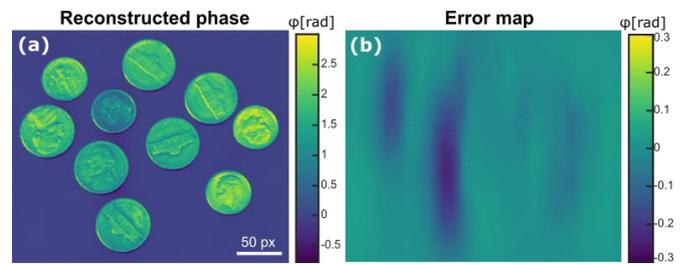

**Figure 6.** Reconstructed object phase and corresponding error map for the case of unequal shear magnitudes.

### 3.2.3 Nonorthogonal shear configuration

We tested OSI-flex using two tilted, non-orthogonal phase-difference maps with shear vectors [2.6, 1.5] (shear magnitude≈3 pixels, orientation≈30°) and [−1.4, 7.9] (shear magnitude≈8 pixels, orientation≈100°), which are mutually tilted by an angle of approximetly 70°. The algorithm used the same parameters as in the previous section: $\eta=0.1$, $\gamma=0$, $\alpha_{TV}=1e^{-8}$, $\alpha_{SP}=2e^{-8}$, 400 iterations. The retrieved phase and error map are shown in Fig. 7. High reconstruction quality was achieved (SSIM=0.9527, RMSE=0.0151), demonstrating OSI-flex's robustness to nonorthogonal shears configuration.

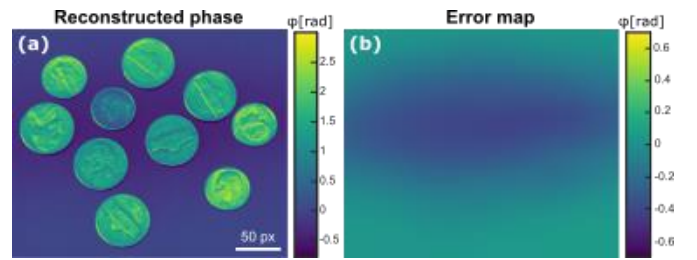

**Figure 7.** Reconstructed object phase and corresponding error map for the case of a pair of tilted, nonorthogonal shears.

### 3.2.4 A single shear configuration

OSI-flex was employed to reconstruction from a single shearing phase measurement. We considered the case of medium (5.3 pixels) vertical shear. The same reconstruction parameters as in the proceeding section was emplyed:, $\eta$=0.1, $\gamma$=0, $\alpha_{TV}$=1e$^{-8}$, $\alpha_{SP}$=2e$^{-8}$, 400 iterations. The reconstructed results are presented in Fig. 8. Since the problem is inherently ill-posed, the reconstruction exhibits some residual errors. Despite these limitations, the essential structural information of the object is well preserved. The obtained image quality measures are: SSIM = 0.7463 and RMSE = 0.1419.

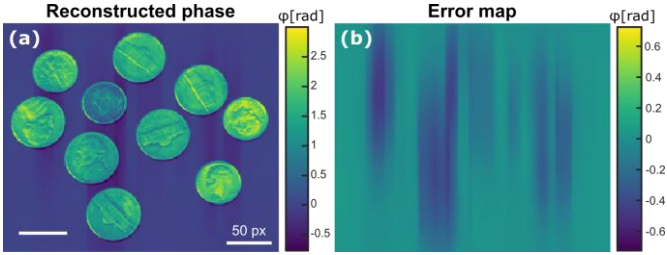

**Figure 8.** Reconstructed object phase and corresponding error map for the case of a single SI measurement.

### 3.2.5 Joint optimization of shear vectors

Accurate shear estimation is a major practical challenge in most optical systems and typically requires calibration with a test object of known phase delay. In this section, we investigate the influence of shear errors on phase restoration and evaluate the fine-tuning capability of OSI-flex.

In this test, we considered two orthogonal shears of medium magnitude (5.3 pixels) but assumed that they were underestimated by 1 pixel, thus we used 4.3 pixels shear value as input to the algorithm. Figures 9(a)–(b) show the OSI-flex reconstruction without shear fine-tuning. The shear error caused significant inaccuracies in the retrieved phase, with underestimation of the shear resulting in overestimated phase values (opposite effect was observed for overestimated shears magnitude). Shear errors also introduced ringing artifacts (Fig. 9(b)).

Importantly, enabling the shear fine-tuning option in OSI-flex effectively mitigated these errors, as shown in Figures 9(c)-(d). To address large shear errors, strong SP regularization was employed to stabilize the optimization process. The algorithm was run with the following parameters: $\eta$=0.1, $\gamma$=1e$^{-4}$, $\alpha_{TV}$=1e$^{-8}$, $\alpha_{SP}$=1e$^{-7}$, 400 iterations. For these settings, the computational time was approximately 18 s, representing a moderate 8% increase due to enabling shear estimation.

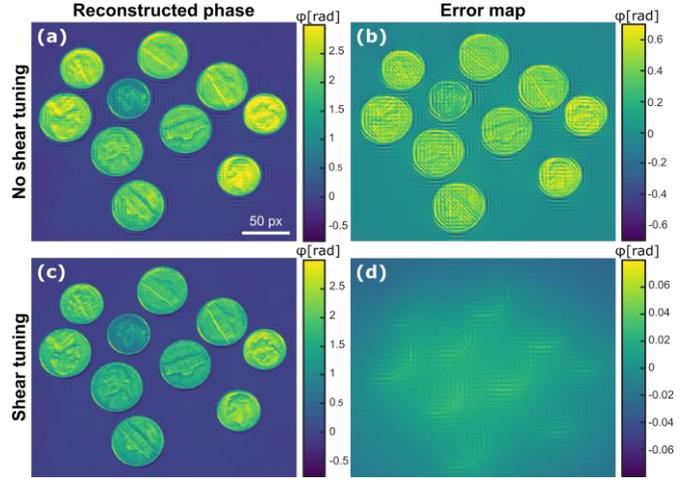

**Figure 9.** Reconstructed object phases and corresponding error maps for the case of a 1-pixel shear error: (a)-(b) without shear fine-tuning, (c)-(d) with shear fine-tuning

**Table 1.** Metrics of reconstructed phases.

| Shear error [px] | | 0.1 | 1 | 2 |
|---|---|---|---|---|
| Without finetuning | SSIM | 0.961 | 0.763 | 0.532 |
| | RMSE | 0.022 | 0.181 | 0.326 |
| With finetuning | SSIM | 0.967 | 0.967 | 0.966 |
| | RMSE | 0.012 | 0.012 | 0.013 |

We repeated the test for relatively small (0.1-pixel) and large (2-pixel) shear errors. The image quality metrics of the retrieved phases are summarized in Table 2. Even a 0.1-pixel error caused a noticeable increase in RMSE, from 0.012 in the ideal case (no shear error) to 0.022. The 2-pixel error resulted in severe deterioration of the reconstruction (RMSE = 0.326, SSIM = 0.532). Nevertheless, in all cases, OSI-flex enabled high-quality phase restoration, with SSIM values above 0.96 and RMSE not exceeding 0.013. For the context, without the shear error, OSI-flex yielded SSIM = 0.967 and RMSE = 0.012.

### 3.3 Experimental results

In this section, we evaluate OSI-flex using experimental data acquired with our SI system (Sec. 2.1). We first consider pairs of SI measurements obtained with approximately orthogonal shear vectors. In all investigated cases, we applied identical optimization parameters ($\eta$=0.1, $\alpha_{TV}$=1e$^{-8}$, $\alpha_{SP}$=1e$^{-8}$, 400 iterations), consistent with those used in the corresponding simulation case (Sec. 3.2.1). For experimental data, where shear vectors are often not known with sufficient accuracy, we employed the shear fine-tuning feature of OSI-flex and set $\gamma = 1e^{-4}$. The initial shear vectors were estimated through visual inspection.

### 3.3.1 Calibrated phase target

First, we evaluate OSI-flex on phase resolution test target (Lyncée Tec), made by etching $125 \pm 5$ nm deep structure in Boroflat 33 glass, corresponding to 0.7039 radians phase delay in our experiment. The two shearing phase differences, which serve as the input to OSI-flex, are shown in Fig. 10 (a),(b). The initial shear vectors were [6, 2] and [2, −7] pixels. The resulting OSI-flex phase reconstruction is presented in Fig. 10 (c). The retrieved phase delay (Fig. 10(d)) agrees well with the theoretical value (blue dotted line). It is also worth noting the uniform background. Importantly, OSI-flex's ability to recover the low spatial frequency band is a significant advantage, as this aspect is typically challenging for differential-based imaging.

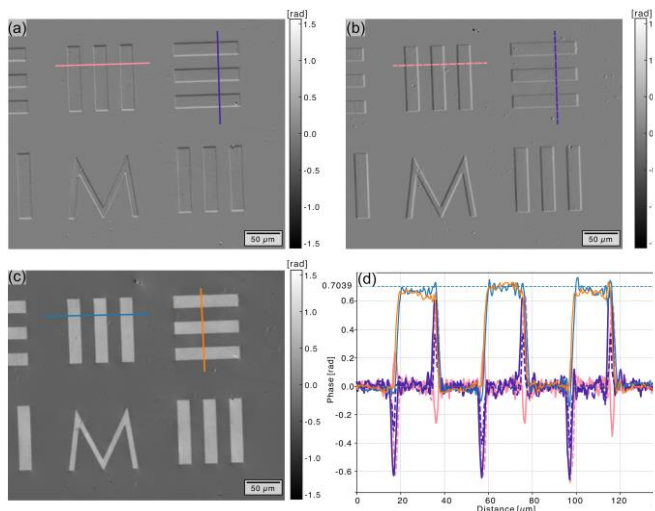

**Figure 10.** Calibration phase target: (a), (b) sheared phase differences, (c) reconstructed object phase, and (d) cross-sections through the data in (a)–(c). The blue dashed line in (d) marks the correct height of the phase resolution target (0.7039 rad).

### 3.3.2 Cell phantom

The next sample is a 3D-printed phantom that mimics the experimentally measured phase distributions of HeLa cells. These cells were arranged in a cluster with relatively high confluence, which is typical for cell culture. Such a sample is challenging for SI object phase reconstruction algorithms as it contains a wide range of spatial frequencies. Details on the design and fabrication of such sample can be found in [80].

The shearing phase data obtained for the cell phantom are shown in Fig. 11(a), (b). The initial shear guesses were [−3.5, −4.5] and [−4.5, 2.0] pixels. The OSI-flex reconstruction is displayed in Fig. 11(c). The retrieved image exhibits the expected uniform background, with no increase in noise level nor error propagation, which are common in other numerical integration algorithms.

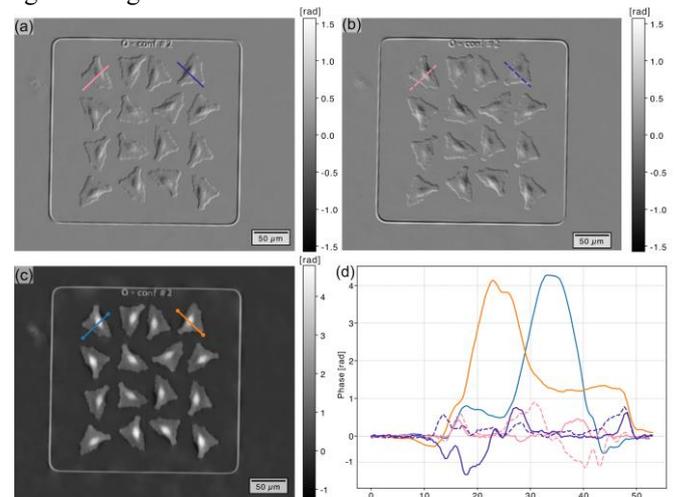

**Figure 11.** Printed cells phantom: (a), (b) sheared phase differences, (c) reconstructed object phase, and (d) cross-sections through the data in (a)–(c).

### 3.3.3 Thyroid smear

Lastly, we consider the measurement of follicular thyroid cells as an example of a biological specimen. Cytological material was obtained from thyroid nodules via fine needle aspiration (FNA) under ultrasonographic guidance. A 25-27G biopsy needle was inserted into the lesion, and multiple passes with varying trajectories were made to sample different areas. The aspirated material was evenly spread on glass slides (cell layer 2-5 µm), fixed in an aerosol-based fixative (Cytofix) at room temperature (20-25 °C), and routinely stained with hematoxylin and eosin (HE).

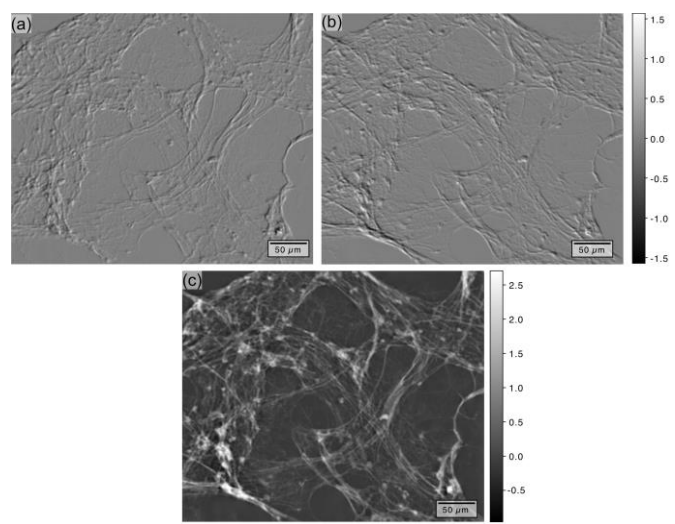

**Figure 12.** Thyroid smear: (a), (b) sheared phase differences, (c) reconstructed object phase.

The sheared phase-difference measurements obtained from the thyroid sample are shown in Fig. 12(a) and (b). The initial shear estimates were [−3.5, −4.5] and [−4.5, 2.0] pixels. The OSI-flex reconstruction is presented in Fig. 12(c). The reconstructed image reveals a complex structure rich in high-spatial-frequency details, showing thyrocytes both in clusters and dispersed throughout the background, along with colloid strands. Notably, no halo effect is observed, indicating effective recovery of the low-spatial-frequency components.

### *3.3.4 Phase reconstruction from a single shearing data*

In this section, we investigate the possibility of object phase retrieval from a single sheared phase-difference measurement. For this purpose we selected one phase-difference image from the pairs presented in Secs. 3.3.1–3.3.3. For OSI-flex reconstruction, we used the same algorithm parameters as in the relevant simulation section (Sec. 3.2.4): $\eta=0.1$, $\alpha_{TV}=1e^{-8}$, $\alpha_{SP}=2e^{-8}$, 400 iterations. These parameters are consistent with those used in the previous experimental sections (Secs. 3.2.1–3.3.2), with the exception that the sign-penalty coefficient $\alpha_{SP}$ was doubled to compensate for the missing information about the phase derivative in the orthogonal direction.

Additionally, because the single-measurement case is ill-posed, we simplified the OSI-flex task by not fine-tuning the shear vector (thus $\gamma=0$). Instead we assumed known shear value. Specifically, we used the shear vector estimated by OSI-flex when applied to the pair of SI measurements (see Secs. 3.3.1–3.3.3).

The results of the reconstruction are presented in Fig. 13. It can be seen that the essential information content of the object phase distribution was successfully retrieved. While the reconstruction quality is reduced and characteristic stripe artifacts along the shear direction are visible, the overall performance remains encouraging and demonstrates the method's effectiveness in the challenging conditions of a single-shear measurement.

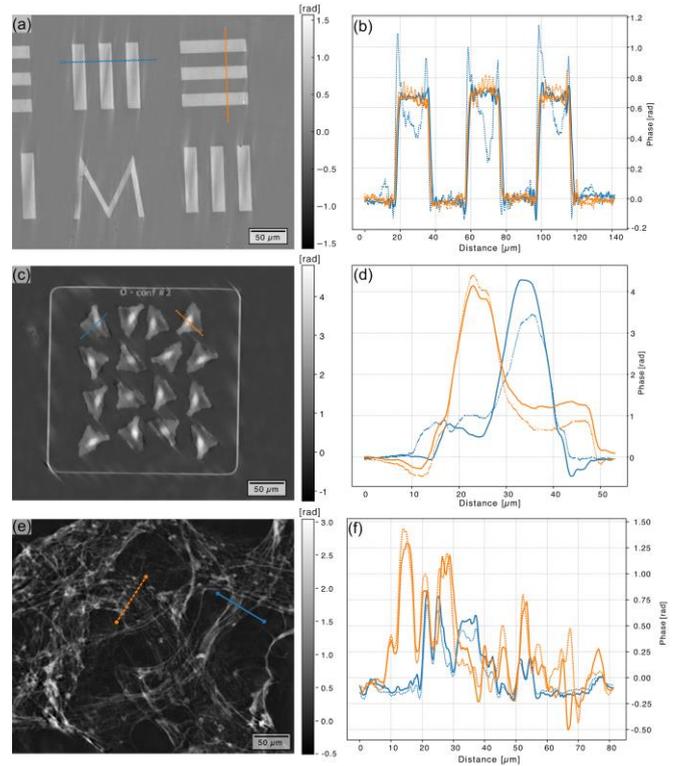

**Figure 13.** Object phase retrieval using a single shear: reconstructed phase (a, c, e) and corresponding cross-sections (b, d, f), including comparison with relevant phase reconstructions obtained using pairs of shearing phase differences (Figs 10(c), 11(c), and 12(c), respectively). The results were obtained for the calibrated phase object (a, b), cell phantom (c, d), and thyroid smear (e, f).

## 4. Discussion

The proposed OSI-flex algorithm takes advantage of modern optimization tools such as automatic differentiation and the ADAM optimizer. Automatic differentiation ensures accurate and efficient computation of gradients. Unlike numerical differentiation, automatic differentiation avoids discretization errors and instability associated with finite-difference schemes. Compared with symbolic differentiation, it scales better to complex, high-dimensional problems and is straightforward to implement within existing optimization frameworks. Moreover, software utilizing automatic differentiation can be easily modified, for example, to accommodate other SI configurations (radial or azimuthal) beyond the lateral SI discussed in this paper.

ADAM provides significant benefits for pure cost-function minimization: it adaptively adjusts learning rates for each parameter, accelerates convergence, and improves stability compared to standard gradient descent. Together, automatic differentiation and ADAM enable fast, robust, and precise optimization, which is essential for large-scale inverse imaging problems.

OSI-flex is highly flexible with respect to the number, magnitudes, and directions of applied shear vectors. Notably, other exisiting solutions for object phase reconstrcution in SI impose various strict restritions on the shear vectors, e.g. they require two shears in perfectly orthogonal *x*- and *y*- directions [21,25,26,49,50,52,53,61,64,66], multiple SI measurements in diversfied directions [30], small shears magitudes [21,52] or even shears of exactly one-pixel magnitudes [22,23,65]. Contrary, OSI-flex is very flexible in terms of shearing settings. As shown in Sec. 3.2.1, our method can handle very small (subpixel) shears, where the shearing phase corresponds to the derivative of the object phase. It can also be applied in realistic experimental conditions where generating very small shear is not possible (medium shear case). Notably, this applies to our experimental system, where the minimal achievable shear is limited by the smallest distance that can be introduced between the polarization diffraction gratings, constrained by their housing. Our simulations further demonstrated that OSI-flex remains effective for large shears, extending to several tens of pixels. In this way, OSI-flex overcomes limitations [81–83] of total-shear SI systems, whose field of view is restricted by the object beam replica and which are often unsuitable for confluent objects.

Regarding the mutual directions of the shears, the optimal condition for phase recovery with OSI-flex is the use of two roughly orthogonal shear directions. Importantly, these directions do not need to be purely vertical and horizontal—tilted directions are equally permissible (see experimental cases in Sec. 3.3). Hower, the phase recovery with OSI-flex is also possible with nonorthogonal shear vectors and even with a single SI measurement. In these cases, a higher amount of regularization is required, and the quality of phase recovery is reduced. Nevertheless, the ability to recover information-bearing phase image from a single shear is particularly important, as it enables monitoring of dynamic processes.

It is important to note that we tested our algorithm on a diverse set of samples containing both low- and high-frequency content. Using the calibrated phase target, we demonstrated the quantitative capability of the OSI-flex algorithm. The presented results show reliable retrieval of both fine details and slowly varying object features. Notably, the same optimization parameters were applied across all measurement cases under analogous SI conditions (one set of parameters for paired SI measurements and another for single SI measurements), demonstrating the robustness of the OSI-flex algorithm.

While the applied SP+TV regularization is best suited for piecewise-constant samples, such as the phase resolution test target in Sec. 3.3.1, it also perform well for gradient-dominated objects. This is because only a small amount of regularization was used—sufficient to stabilize the optimization procedure without oversmoothing the reconstructions.

Lastly, our simulations showed that a shear error as small as 0.1 pixel can lead to noticeable phase errors, whereas, according to Ref. [30], an alternative correlation-based method can detect shear with an accuracy of up to 0.5 pixels. Moreover, the cross-correlation method may be unreliable for small shear magnitudes due to the close proximity of the correlation peaks. In Sec. 3.2.5, we demonstrated that the shear fine-tuning feature of OSI-flex effectively minimizes the degradation of phase quality caused by incorrect assumptions about the shear value. Joint estimation of phase and shear is crucial for reliable processing of SI data in practical experimental settings.

OSI-flex fills a critical gap in high-quality, open-source algorithms for phase recovery in SI. Unlike existing software [66], which is restricted to ideally orthogonal, *x*- and *y*-oriented, small and known shear magnitudes, OSI-flex is highly flexible and can be applied to a wide range of SI configurations. The algorithm and accompanying test data are publicly available, and its implementation using modern tool—TensorFlow—ensures efficient, robust performance. We anticipate that OSI-flex will accelerate the development and adoption of SI and other differentiation-based imaging techniques, e.g, stereo deflectometry [55].

## 5. Conclusions

We introduce OSI-flex, a robust and versatile algorithm for phase recovery in SI. By combining automatic differentiation with the ADAM optimizer, OSI-flex delivers fast, accurate, and stable gradient-based optimization. Unlike conventional approaches, OSI-flex is flexible with respect to applied shear numbers, magnitudes, and directions. Moreover, OSI-flex does not rely solely on the nominal shear values, as it can jointly optimize both phase and shear parameters. This capability ensures high-quality reconstructions in realistic experimental conditions where the shear vectors are not known with sufficient accuracy.

Extensive testing with simulations and experimental measurements, including a calibrated phase resolution test target, confirms the quantitative accuracy and reliability of the algorithm. OSI-flex maintains consistent performance across samples with varying structural complexity, demonstrating accurate recovery of both low and high spatial frequencies and strong robustness against common numerical integration issues, such as error propagation and noise amplification. Notably, OSI-flex can reconstruct object information even from a single shear measurement, enabling dynamic process monitoring.

As an open-source tool, OSI-flex fills a critical gap in existing phase-recovery software and is expected to accelerate the development and adoption of SI and other differentiation-based phase imaging techniques.

**Acknowledgements**

This work was funded by the National Science Center Poland (2024/55/D/ST7/02792). This work was partially funded by the project no. WPC3/2022/47/INTENCITY/2024 funded by the National Centre for Research and Development (NCBR) under the 3rd Polish-Chinese/Chinese-Polish Joint Research Call (2022). The study was carried out on devices co-funded by the Warsaw University of Technology within the Excellence Initiative: Research University (IDUB) program.

**Authors contributions**

JW: Conceptualization, Data Curation, Formal Analysis, Investigation, Methodology, Software, Supervision, Validation, Visualization, Writing – Original Draft. DS: Formal Analysis, Validation, Writing – Original Draft, Writing – Review & Editing. MH: Data Curation, Investigation. ML: Resources. MZ: Resources. MT: Funding Acquisition, Project Administration, Resources, Supervision, Writing – Review & Editing. PZ: Conceptualization, Data Curation, Funding Acquisition, Investigation, Methodology, Project Administration, Supervision, Visualization, Writing – Original Draft, Writing – Review & Editing.

**Data and code availability**

The OSI-flex software and datasets used during the current study are available at [73] and [84], repsectively, available upon publication.

**References**

[1]　Popescu G 2011 *Quantitative Phase Imaging of Cells and Tissues* (McGraw-Hill Education)

[2]　Park Y, Depeursinge C and Popescu G 2018 Quantitative phase imaging in biomedicine *Nat. Photonics* **12** 578–89

[3]　Anand A, Chhaniwal V and Javidi B 2018 Tutorial: Common path self-referencing digital holographic microscopy *APL Photonics* **3** 071101

[4]　Zhang J, Dai S, Ma C, Xi T, Di J and Zhao J 2021 A review of common-path off-axis digital holography: towards high stable optical instrument manufacturing *Light Adv. Manuf.* **2** 333–49

[5]　Zdańkowski P, Winnik J, Patorski K, Gocłowski P, Ziemczonok M, Józwik M, Kujawińska M and Trusiak M 2021 Common-path intrinsically achromatic optical diffraction tomography *Biomed. Opt. Express* **12** 4219–34

[6]　Bhaduri B, Edwards C, Pham H, Zhou R, Nguyen T H, Goddard L L and Popescu G 2014 Diffraction phase microscopy: principles and applications in materials and life sciences *Adv. Opt. Photonics* **6** 57–119

[7]　Velghe S, Primot J, Guérineau N, Cohen M and Wattellier B 2005 Wave-front reconstruction from multidirectional phase derivatives generated by multilateral shearing interferometers *Opt. Lett.* **30** 245–7

[8]　Popescu G, Ikeda T, Dasari R R and Feld M S 2006 Diffraction phase microscopy for quantifying cell structure and dynamics *Opt. Lett.* **31** 775–7

[9]　Joo K-N and Park H M 2023 Shearing Interferometry: Recent Research Trends and Applications *Curr. Opt. Photonics* **7** 325–36

[10]　Tayal S, Usmani K, Singh V, Dubey V and Singh Mehta D 2019 Speckle-free quantitative phase and amplitude imaging using common-path lateral shearing interference microscope with pseudo-thermal light source illumination *Optik* **180** 991–6

[11]　Tang X, Cao Z, Wang Z, Xie H and Xu L 2021 Retrieval of Phase and Temperature Distributions in Axisymmetric Flames From Phase-Modulated Large Lateral Shearing Interferogram *IEEE Trans. Instrum. Meas.* **70** 1–12

[12]　Lee K and Park Y 2014 Quantitative phase imaging unit *Opt. Lett.* **39** 3630–3

[13]　Baek Y, Lee K, Yoon J, Kim K and Park Y 2016 White-light quantitative phase imaging unit *Opt. Express* **24** 9308–15

[14]　Zdańkowski P, Winnik J, Rogalski M, Marzejon M J, Wdowiak E, Dudka W, Józwik M and Trusiak M 2025 Polarization Gratings Aided Common-Path Hilbert Holotomography for Label-Free Lipid Droplets Content Assay *Laser Photonics Rev.* **19** 2401474

[15]　Edwards C, Zhou R, Hwang S-W, McKeown S J, Wang K, Bhaduri B, Ganti R, Yunker P J, Yodh A G, Rogers J A, Goddard L L and Popescu G 2014 Diffraction phase microscopy: monitoring nanoscale dynamics in materials science [Invited] *Appl. Opt.* **53** G33–43

[16]　Tang Z, Winnik J and Hennelly B M 2025 Optical diffraction tomography using a self-reference module *Biomed. Opt. Express* **16** 57–67

[17]　Allen R D, David G B and Nomarski G 1969 The zeiss-Nomarski differential interference equipment for transmitted-light microscopy *Z. Wiss. Mikrosk. Mikrosk. Tech.* **69** 193–221

[18] Murty M V R K 1964 The Use of a Single Plane Parallel Plate as a Lateral Shearing Interferometer with a Visible Gas Laser Source *Appl. Opt.* **3** 531–4

[19] Rimmer M P and Wyant J C 1975 Evaluation of Large Aberrations Using a Lateral-Shear Interferometer Having Variable Shear *Appl. Opt.* **14** 142–50

[20] Zhu Y, Tian A, Wang H, Liu B and Wang K 2025 A novel multiple directional shearing interferometry system with synchronous polarization phase shifting *Sci. Rep.* **15** 2775

[21] Ghim Y-S, Rhee H-G, Davies A, Yang H-S and Lee Y-W 2014 3D surface mapping of freeform optics using wavelength scanning lateral shearing interferometry *Opt. Express* **22** 5098–105

[22] Rubio-Oliver R, Micó V, Zalevsky Z, García J and Angel Picazo-Bueno J 2024 Quantitative phase imaging by automated Cepstrum-based interferometric microscopy (CIM) *Opt. Laser Technol.* **177** 111121

[23] Rubio-Oliver R, García J, Zalevsky Z, Picazo-Bueno J Á and Micó V 2024 Cepstrum-based interferometric microscopy (CIM) for quantitative phase imaging *Opt. Laser Technol.* **174** 110626

[24] Ma C, Li Y, Zhang J, Li P, Xi T, Di J and Zhao J 2017 Lateral shearing common-path digital holographic microscopy based on a slightly trapezoid Sagnac interferometer *Opt. Express* **25** 13659–67

[25] Bon P, Maucort G, Wattellier B and Monneret S 2009 Quadriwave lateral shearing interferometry for quantitative phase microscopy of living cells *Opt. Express* **17** 13080–94

[26] Xie J, Xie H, Kong C Z and Ling T 2024 Quadri-wave lateral shearing interferometry: a versatile tool for quantitative phase imaging *JOSA A* **41** C137–56

[27] Baffou G 2023 Wavefront Microscopy Using Quadriwave Lateral Shearing Interferometry: From Bioimaging to Nanophotonics *ACS Photonics* **10** 322–39

[28] Han Z-G, Meng L-Q, Huang Z-Q, Shen H, Chen L and Zhu R-H 2017 Determination of the laser beam quality factor ($M^2$) by stitching quadriwave lateral shearing interferograms with different exposures *Appl. Opt.* **56** 7596–603

[29] Marthy B, Bénéfice M and Baffou G 2024 Single-shot quantitative phase-fluorescence imaging using cross-grating wavefront microscopy *Sci. Rep.* **14** 2142

[30] Shanmugam P, Light A, Turley A and Falaggis K 2022 Variable shearing holography with applications to phase imaging and metrology *Light Adv. Manuf.* **3** 193–210

[31] Winnik J, Zdankowski P, Stefaniuk M, Ahmad A, Zuo C, Ahluwalia B S and Trusiak M 2024 Gradient Optical Diffraction Tomography *Prepr. ArXiv241108423*

[32] Nguyen M T, Park H-M, Joo K-N and Ghim Y-S 2025 Deep learning-based single-shot lateral shearing interferometry *Opt. Lasers Eng.* **191** 109010

[33] Pant K K, Burada D R, Bichra M, Ghosh A, Khan G S, Sinzinger S and Shakher C 2018 Weighted spline based integration for reconstruction of freeform wavefront *Appl. Opt.* **57** 1100–9

[34] Yao F, Qian C, Jiasong S, Zuxin Z, Linpeng L and Chao Z 2019 Review of the development of differential phase contrast microscopy *Infrared Laser Eng.* **48** 0603014

[35] Nguyen T H, Kandel M E, Rubessa M, Wheeler M B and Popescu G 2017 Gradient light interference microscopy for 3D imaging of unlabeled specimens *Nat. Commun.* **8** 210

[36] Kandel M E, Hu C, Naseri Kouzehgarani G, Min E, Sullivan K M, Kong H, Li J M, Robson D N, Gillette M U, Best-Popescu C and Popescu G 2019 Epi-illumination gradient light interference microscopy for imaging opaque structures *Nat. Commun.* **10** 4691

[37] Zhang J, Sarollahi M, Luckhart S, Harrison M J and Vasdekis A E 2024 Quantitative phase imaging by gradient retardance optical microscopy *Sci. Rep.* **14** 9754

[38] Kwon H, Arbabi E, Kamali S M, Faraji-Dana M and Faraon A 2020 Single-shot quantitative phase gradient microscopy using a system of multifunctional metasurfaces *Nat. Photonics* **14** 109–14

[39] Wang X, Wang H, Wang J, Liu X, Hao H, Tan Y S, Zhang Y, Zhang H, Ding X, Zhao W, Wang Y, Lu Z, Liu J, Yang J K W, Tan J, Li H, Qiu C-W, Hu G and Ding X 2023 Single-shot isotropic differential interference contrast microscopy *Nat. Commun.* **14** 2063

[40] Ledwig P and Robles F E 2021 Quantitative 3D refractive index tomography of opaque samples in epi-mode *Optica* **8** 6–14

[41] Casteleiro Costa P, Wang B, Elizabeth Serafini C, Bowles-Welch A, Yeago C, Roy K and Robles F E 2022 Functional imaging with dynamic quantitative oblique

back-illumination microscopy *J. Biomed. Opt.* **27** 066502

[42] Strasser F, Ritsch-Marte M and Bernet S 2025 Quantitative phase imaging with optical differentiation by spatially variable amplitude filters *Opt. Lett.* **50** 353–6

[43] Valera J D and Jones J D C 1994 Phase stepping in fiber-based speckle shearing interferometry *Opt. Lett.* **19** 1161–3

[44] Patorski K 1988 Shearing interferometry and the moire method for shear strain determination *Appl. Opt.* **27** 3567–72

[45] Assa A, Politch J and Betser A A 1979 Slope and curvature measurement by a double-frequency-grating shearing interferometer *Exp. Mech.* **19** 129–37

[46] Yokozeki S and Ohnishi K 1975 Spherical Aberration Measurement with Shearing Interferometer Using Fourier Imaging and Moiré Method *Appl. Opt.* **14** 623–7

[47] Yatagai T and Kanou T 1984 Aspherical Surface Testing with Shearing Interferometer Using Fringe Scanning Detection Method *Opt. Eng.* **23** 234357

[48] Tian X, Itoh M and Yatagai T 1995 Simple algorithm for large-grid phase reconstruction of lateral-shearing interferometry *Appl. Opt.* **34** 7213–20

[49] Okuda S, Nomura T, Kamiya K, Miyashiro H, Yoshikawa K and Tashiro H 2000 High-precision analysis of a lateral shearing interferogram by use of the integration method and polynomials *Appl. Opt.* **39** 5179–86

[50] Liang P, Ding J, Jin Z, Guo C-S and Wang H 2006 Two-dimensional wave-front reconstruction from lateral shearing interferograms *Opt. Express* **14** 625–34

[51] Bon P, Monneret S and Wattellier B 2012 Noniterative boundary-artifact-free wavefront reconstruction from its derivatives *Appl. Opt.* **51** 5698–704

[52] Huang L, Idir M, Zuo C, Kaznatcheev K, Zhou L and Asundi A 2015 Comparison of two-dimensional integration methods for shape reconstruction from gradient data *Opt. Lasers Eng.* **64** 1–11

[53] Ling T, Jiang J, Zhang R and Yang Y 2017 Quadriwave lateral shearing interferometric microscopy with wideband sensitivity enhancement for quantitative phase imaging in real time *Sci. Rep.* **7** 9

[54] Konijnenberg A P, Beurs A C C de, Jansen G S M, Urbach H P, Witte S and Coene W M J 2020 Phase retrieval algorithms for lensless imaging using diffractive shearing interferometry *JOSA A* **37** 914–24

[55] Ren H, Gao F and Jiang X 2016 Least-squares method for data reconstruction from gradient data in deflectometry *Appl. Opt.* **55** 6052–9

[56] Harbers G, Kunst P J and Leibbrandt G W 1996 Analysis of lateral shearing interferograms by use of Zernike polynomials *Appl. Opt.* **35** 6162–72

[57] Mochi I and Goldberg K A 2015 Modal wavefront reconstruction from its gradient *Appl. Opt.* **54** 3780

[58] van Brug H 1997 Zernike polynomials as a basis for wave-front fitting in lateral shearing interferometry *Appl. Opt.* **36** 2788–90

[59] Dai F, Tang F, Wang X, Sasaki O and Feng P 2012 Modal wavefront reconstruction based on Zernike polynomials for lateral shearing interferometry: comparisons of existing algorithms *Appl. Opt.* **51** 5028–37

[60] Southwell W H 1980 Wave-front estimation from wave-front slope measurements *JOSA Vol 70 Issue 8 Pp 998-1006*

[61] Nguyen V-H-L, Rhee H-G and Ghim Y-S 2022 Improved Iterative Method for Wavefront Reconstruction from Derivatives in Grid Geometry *Curr. Opt. Photonics* **6** 1–9

[62] Li G, Li Y, Liu K, Ma X and Wang H 2013 Improving wavefront reconstruction accuracy by using integration equations with higher-order truncation errors in the Southwell geometry *JOSA A* **30** 1448–59

[63] Ji Z, Zhang X, Zheng Z, Li Y and Chang J 2020 Algorithm based on the optimal block zonal strategy for fast wavefront reconstruction *Appl. Opt.* **59** 1383–96

[64] Huang L, Xue J, Gao B, Zuo C and Idir M 2017 Spline based least squares integration for two-dimensional shape or wavefront reconstruction *Opt. Lasers Eng.* **91** 221–6

[65] Liu X, Gao Y and Chang M 2009 A partial differential equation algorithm for wavefront reconstruction in lateral shearing interferometry *J. Opt. Pure Appl. Opt.* **11** 045702

[66] D'Errico J 2025 Inverse (integrated) gradient https://www.mathworks.com/matlabcentral/fileexchange/9734-inverse-integrated-gradient

[67] de Groot P 2011 Phase Shifting Interferometry *Optical Measurement of Surface Topography* ed R Leach (Berlin, Heidelberg: Springer) pp 167–86

[68] Takeda M, Ina H and Kobayashi S 1982 Fourier-transform method of fringe-pattern analysis for computer-based topography and interferometry *JOSA* **72** 156–60

[69] Trusiak M, Cywińska M, Micó V, Picazo-Bueno J Á, Zuo C, Zdańkowski P and Patorski K 2020 Variational Hilbert Quantitative Phase Imaging *Sci. Rep.* **10** 13955

[70] Kingma D P and Ba J 2017 Adam: A Method for Stochastic Optimization

[71] Bian D, Kim D, Kim B, Yu L, Joo K-N and Kim S-W 2020 Diverging cyclic radial shearing interferometry for single-shot wavefront sensing *Appl. Opt.* **59** 9067–74

[72] Willard B C 1993 Rotational Shearing Interferometer *Appl. Opt.* **32** 7118–9

[73] Winnik J QCI-LAB/OSI-flex: Flexible, optimization-based algorithm for joint phase and shear estimation from shearing interferometry data https://github.com/QCI-LAB/OSI-flex

[74] Abadi M, Agarwal A, Barham P, Brevdo E, Chen Z, Citro C, Corrado G s, Davis A, Dean J, Devin M, Ghemawat S, Goodfellow I, Harp A, Irving G, Isard M, Jia Y, Jozefowicz R, Kaiser L, Kudlur M and Zheng X 2016 TensorFlow: Large-Scale Machine Learning on Heterogeneous Distributed Systems

[75] Abadi M, Barham P, Chen J, Chen Z, Davis A, Dean J, Devin M, Ghemawat S, Irving G, Isard M, Kudlur M, Levenberg J, Monga R, Moore S, Murray D G, Steiner B, Tucker P, Vasudevan V, Warden P, Wicke M, Yu Y and Zheng X 2016 TensorFlow: a system for large-scale machine learning *Proceedings of the 12th USENIX conference on Operating Systems Design and Implementation* OSDI'16 (USA: USENIX Association) pp 265–83

[76] Duchi J, Hazan E and Singer Y 2011 Adaptive Subgradient Methods for Online Learning and Stochastic Optimization *J. Mach. Learn. Res.* **12** 2121–59

[77] Nocedal J and Wright S J 2006 Conjugate Gradient Methods *Numerical Optimization* (New York, NY: Springer) pp 101–34

[78] Reddi S J, Kale S and Kumar S 2019 On the Convergence of Adam and Beyond *Prepr. ArXiv190409237*

[79] Gao Y and Cao L 2023 Iterative projection meets sparsity regularization: towards practical single-shot quantitative phase imaging with in-line holography *Light Adv. Manuf.* **4** 37–53

[80] Desissaire S, Ziemczonok M, Cantat-Moltrecht T, Kuś A, Godefroy G, Hervé L, Paviolo C, Krauze W, Allier C, Mandula O and Kujawińska M 2025 Bio-inspired 3D-printed phantom: Encoding cellular heterogeneity for characterization of quantitative phase imaging *Measurement* **247** 116765

[81] Mico V, Ferreira C, Zalevsky Z and García J 2014 Spatially-multiplexed interferometric microscopy (SMIM): converting a standard microscope into a holographic one *Opt. Express* **22** 14929–43

[82] Patorski K, Zdańkowski P and Trusiak M 2020 Grating deployed total-shear 3-beam interference microscopy with reduced temporal coherence *Opt. Express* **28** 6893–908

[83] Bai H, Shan M, Zhong Z, Guo L and Zhang Y 2015 Common path interferometer based on the modified Michelson configuration using a reflective grating *Opt. Lasers Eng.* **75**

[84] Winnik J and Zdańkowski P 2025 OSI-flex dataset: sheared phase-difference measurements from experiments and simulations Zenodo https://zenodo.org/records/17119986